\documentclass[prd,aps,preprint,tightenlines,superscriptaddress]{revtex4}

\usepackage{amsmath}
\usepackage{epsfig}

\newcommand{\insertfig}[2]{\mbox{\epsfxsize=#1cm \epsfbox{#2.eps}}}
\preprint{DOE/ER/40762-359} \preprint{MU-PP\#05-006}

\begin{document}

\title{Leptogenesis in Realistic SO(10) Models}
\author{Xiangdong Ji}
\affiliation{Department of Physics, University of Maryland, College
Park, Maryland 20742} \affiliation{Institute for Theoretical
Physics, Peking University, Beijing, China}
\author{Yingchuan Li}
\affiliation{Department of Physics, University of Maryland, College Park,
Maryland 20742}
\author{R. N. Mohapatra}
\affiliation{Department of Physics, University of Maryland, College
Park, Maryland 20742}
\author{S. Nasri}
\affiliation{Department of Physics, University of Maryland, College
Park, Maryland 20742}
\author{Yue Zhang}
\affiliation{Institute for Theoretical Physics, Peking University,
Beijing, China}
\date{\today}

\begin{abstract}
We study the origin of baryonic matter via leptogenesis in realistic
SO(10) models, in particular, in a new lopsided mass matrix model
introduced recently by three of the authors. By introducing simple
CP-violating phases in the mass matrix of the right-handed
neutrinos, the model generates sufficient baryon asymmetry without
fine-tuning. We compare this result with other realistic SO(10)
models.

\end{abstract}

\maketitle

\section{introduction}

One of the most fundamental questions in modern cosmology is where
the baryon number asymmetry in today's Universe comes from. It is an
attractive idea to assume that it arises dynamically in a symmetric
big bang model through C and CP violating processes out of thermal
equilibrium. Of the several proposals that use this scenario,
leptogenesis \cite{Fukugita:1986hr} has emerged as one of the most
interesting scenarios for baryon generation. Here the baryon number
is produced through the so-called sphaleron processes from a
residual lepton number density left over from an early stage of the
universe through lepton-number violating decays of right handed
neutrinos. Such connection is extremely interesting since the same
heavy right-handed neutrinos are also important ingredients for
understanding the small left-handed neutrino masses through the
so-called see-saw mechanism \cite{seesaw}.

Recently much study has been made in the literature about the
feasibility of such a scenario \cite{Buchmuller:2004nz}. Although
one can discuss leptogenesis and neutrino mass texture in models
without introducing quark degrees of freedom, the most interesting
theoretical frameworks involve quark-lepton unification, possibly
under supersymmetric formalism. In addition to connecting the
origin of matter to the big picture of particle physics, the grand
unified theory (GUT) models build in more constraints and have
better predictive power \cite{Mohapatra:2004za}. For various
reasons, SUSY SO(10) has been a favored framework to unify physics
beyond the standard model (for a recent review of SO(10) models,
see \cite{Chen:2003zv}). Depending upon which set of Higgs
multiplets is chosen to break the GUT group and electroweak
symmetry, two classes of SO(10) models are most studied in the
literature: One uses $10_H$, $126_H$, $\overline{126}_H$ and
$120_H$ \cite{Chen:2000fp,GMN,Goh:2003sy,DMM}, and the other uses
$10_H$, $16_H$, $\overline{16}_H$ and $45_H$
\cite{Babu:1998wi,Albright:1998vf,Blazek:1999hz,Ji}. While most of
these models are quite successful in fitting and predicting the
known experimental masses and mixing angles of leptons and quarks,
they predict very different values for the poorly-known neutrino
mixing angle $\theta_{13}$. Quite interestingly, they also paint
different pictures for leptogenesis.

In this paper, we are mostly interested in the leptogenesis in a
lopsided SO(10) mass matrix model proposed recently by three of us
(X.J, Y.L. and R. N. M.) \cite{Ji}. The model is a modification of
the lopsided model originally proposed by Albright, Babu and Barr
\cite{Albright:1998vf}. In the modified version, the right-handed
neutrino mass matrix has a simple diagonal form. The large solar
mixing angle is mainly generated from the neutrino Dirac mass
matrices. The prediction of this model for $\sin\theta_{13}$ is much
larger than the original model. Here we will show that the
leptogenesis picture is also strikingly different. In particular,
enough baryon symmetry can be produced without requiring the two
heavy right-handed neutrino masses, $M_1$ and $M_2$, to be
quasi-degenerate.

The presentation of the paper is as follows. In Section II, we
review the new lopsided SO(10) model. In Section III, we consider in
detail leptogenesis in this model. In Section IV, we make a
comparison of the leptogenesis with other SO(10) models, emphasizing
similarities and differences. We present our conclusions in Section
V.

\section{A New Lopsided SO(10) Model}

A new SUSY SO(10) GUT model with lopsided mass matrices was
introduced in \cite{Ji}. In this section, we briefly describe its
physics. The model was obtained by modifying the right-handed
neutrino, up-quark, and left-handed neutrino mass matrices of the
original lopsided model of Albright and Barr
\cite{Albright:1998vf,Albright:2001LMA,Albright:2000dk}, which we
will describe briefly in Section IV. The lopsidedness built within
the Yukawa couplings between the second and third families
generates, among other interesting physical consequences, the large
atmospheric-neutrino mixing angle $\theta_{23}$ while keeping
$V_{cb}$ in the Cabbibo-Kobayashi-Moskawa (CKM) matrix small. The
modified lopsided model assumes that the right-handed neutrino
Majorana mass matrix $M_R$ has a simple diagonal structure, and
introduces additional off-diagonal couplings in the upper-type-quark
and neutrino Dirac mass matrices to generate $1$-$2$ (solar angle)
rotation. All the fermion masses and mixing angles can be fitted
well in the new model. The mixing angle $\theta_{13}$, however, is
close to the upper limit from the CHOOZ experiment and therefore
definitely within the range of next generation reactor experiments.

We use the convention that Yukawa couplings in the lagrangian appear
as
\begin{equation}
   {\cal L} = - \overline Q_i H Y_{ij} d_{Rj} + ... \ .
\end{equation}
Then the fermion matrix matrices are $M_{ij} = vY_{ij}$ with $v=174$
GeV. Through couplings with a set of Higgs multiplets $10_H$,
$16_H$, $\overline{16}_H$ and $45_H$, the up-type-quark,
down-type-quark, charged-lepton and neutrino Dirac and Majorana mass
matrices in the model of Ref. \cite{Ji} have the following forms,
\begin{eqnarray}
M_u&=&\begin{pmatrix} \eta & 0 & \kappa-\rho/3 \\ 0 & 0 & \omega \\
\kappa+\rho/3 & \omega & 1 \end{pmatrix}\Lambda_U\ ,~~~~
M_{\nu_D}=\begin{pmatrix} \eta & 0 & \kappa+\rho \\ 0 & 0 & \omega \\
\kappa-\rho & \omega & 1 \end{pmatrix}\Lambda_U\ , \nonumber \\
M_d&=&\begin{pmatrix} \eta & \delta & \delta'e^{-i\phi} \\
\delta & 0 & -\epsilon/3 \\ \delta'e^{-i\phi} & \sigma+\epsilon/3
& 1
\end{pmatrix}\Lambda_D \ , ~~~~
M_l=\begin{pmatrix} \eta & \delta & \delta'e^{-i\phi} \\
\delta & 0 & \sigma +\epsilon \\ \delta'e^{-i\phi} & -\epsilon & 1
\end{pmatrix}\Lambda_D \ ,~~ \nonumber \\
M_{\nu_R}&=&\begin{pmatrix} a & 0 & 0 \\
0 & b & 0 \\ 0 & 0 & 1
\end{pmatrix}\Lambda_R \ .
\end{eqnarray}
From the above, we get the Majorana mass matrix of left-handed
neutrinos from the see-saw formula,
$m_{\nu_L}=-M_{\nu_D}M^{-1}_{\nu_R} M_{\nu_D}^T$,
\begin{eqnarray}
m_{\nu_L}=-\begin{pmatrix} \eta^2/a+(\kappa+\rho)^2 &
(\kappa+\rho)\omega & \eta(\kappa-\rho)/a+(\kappa+\rho)
\\ (\kappa+\rho)\omega & \omega^2 & \omega
\\
\eta(\kappa-\rho)/a+(\kappa+\rho) & \omega & 1+
(\kappa-\rho)^2/a+\omega^2/b\end{pmatrix}M^2_U/\Lambda_R \ .
\end{eqnarray}
The parameter $\sigma$ is of order one, signaling the lopsidedness
between the second and third families in $M_d$ and $M_l$. This
feature leads to a large left-handed neutrino mixing in the PMNS
matrix and a small left-handed quark mixing shown in the CKM matrix.
The parameter $\epsilon$ is one order-of-magnitude smaller than
$\sigma$ and generates the hierarchy between the second and third
families. In extending to the first family, $\delta$ and $\delta'$
were introduced into the $M_d$ and $M_l$. Usually a large rotation
in the 1-2 sector is in conflict with the hierarchial property of
the quark masses. However, in the above texture, the $1$-$2$
rotation angle from $M_u$ will be combined with the $1$-$2$ rotation
from $M_d$ to obtain the Cabibbo angle $\theta^{c}$, and thus the
constraint from the up-type quark spectrum is avoided. The first two
families in the $M_u$ and $M_{\nu_L}$ are not coupled to each other
directly but through couplings with the third family. The rotations
in 1-2 sector generated for left-handed up-type quarks and neutrinos
are proportional to the ratios $\gamma\equiv(\kappa-\rho/3)/\omega$
and $\gamma'\equiv(\kappa+\rho)/\omega$, respectively.

The procedure to fit various parameters to experimental data is as
follows. First, the up-type quark and lepton spectra and the
parameters in the CKM matrix are used to determine 10 parameters:
$\sigma$, $\epsilon$, $\delta$, $\delta'$, $\phi$, $\omega$,
$\gamma$, $\eta$, $\Lambda_U$ and $\Lambda_D$. The best fit yields
$\sigma$ and $\epsilon$ approximately the same as those in the
original lopsided model, and thus the successful prediction for the
mass ratios $m^0_{\mu}/m^0_{\tau}$ and $m^0_s/m^0_b$ are kept. The
down-type quark mass spectrum comes out as predictions. The present
model constrains the neutrino mass spectrum as hierarchial. The mass
difference $\Delta m^2_{\nu 12}$ is used to fix the right-handed
neutrino mass scale $\Lambda_R$.

We summarize our input and detailed fits as follows. For CKM
matrix elements, we take $|V_{us}|=0.224$, $|V_{ub}|=0.0037$,
$|V_{cb}|=0.042$, and $\delta_{CP}=60^\circ$ as inputs at the
electroweak scale. With a running factor of $0.8853$ for
$|V_{ub}|$, and $|V_{cb}|$ taken into account, we have
$|V^0_{ub}|=0.0033$ and $|V^0_{cb}|=0.037$ at the GUT scale. For
charged lepton masses and up quark masses, we take the values at
the GUT scale corresponding to $\rm{tan}\beta=10$ from Ref.
\cite{das}. For neutrino oscillation data, we take the
solar-neutrino angle to be $\theta_{\rm solar}=32.5^\circ$ and
mass square differences as $\Delta
m^2_{\nu12}=7.9\times10^{-5}\rm{eV}^2$ and $\Delta
m^2_{\nu23}=2.4\times10^{-3}\rm{eV}^2$. The result for the 12
fitted parameters is
\begin{eqnarray} \sigma&=&1.83\ , ~~\epsilon=0.1446\ , ~~
\delta=0.01\ , \nonumber \\
\delta'&=&0.014\ ,~~\phi=27.9^\circ \ , ~~\eta=1.02\times10^{-5}\ , \nonumber \\
\omega&=&-0.0466\ ,~~\rho=0.0092\ , ~~\kappa=0.0191\ , \nonumber \\
M_{U}&=&82.2~ {\rm GeV}\ ,~~ M_{D}=583.5~{\rm MeV}\ ,
~~\Lambda_R=1.85\times10^{13}~{\rm GeV}\ .
\end{eqnarray}
There is a combined constraint on $a$ and $b$, and thus the
right-handed Majorana mass spectrum is not well determined. As
examples, if $a=b$, $a=-2.039\times 10^{-3}$; and if $a=1$,
$b=-1.951\times 10^{-3}$.

In our previous paper, we have taken $a=b$ and real. The results
for the down-type quark masses and right-handed Majorana neutrino
masses are as follows,
\begin{eqnarray}
&& m^0_d=1.08~{\rm MeV}\ , ~~ m^0_s=25.97~{\rm MeV}\ , ~~
m^0_b=1.242~
{\rm GeV}\ , \nonumber \\
&&M_1=3.77\times10^{10}{\rm GeV}\ , ~~ M_2=3.77\times10^{10}{\rm
GeV}\ , ~~ M_3=1.85\times10^{13}{\rm GeV} \ .
\end{eqnarray}
The predictions for the mixing angles in the PMNS matrix are,
\begin{equation}
\sin^2\theta_{\rm atm}=0.49\ , ~~ \sin^22\theta_{13}=0.074 \ .
\end{equation}
If one releases the best-fit value of $\Delta m^2_{\nu12}$ and
$\Delta m^2_{\nu23}$ and impose only the $3\sigma$ constraint as
$7.1\times10^{-5}\rm{eV}^2\leq\Delta
m^2_{\nu12}\leq8.9\times10^{-5}\rm{eV}^2$ and
$1.4\times10^{-3}\rm{eV}^2\leq\Delta
m^2_{\nu23}\leq3.3\times10^{-3}\rm{eV}^2$, one would obtain,
$0.44\leq\sin^2\theta_{\rm atm}\leq0.52$ which is well within the
$1\sigma$ limit, and $0.055\leq\sin^22\theta_{13}\leq0.110$ which,
as a whole region, lies in the scope of the next generation of
reactor experiments.

\section{Leptogenesis}

The baryon asymmetry of the universe is customarily defined as the
ratio of the baryon density and the photon density after
recombination, and has been measured to very good precision from the
WMAP experiment \cite{Spergel:2006hy}:
\begin{equation}
  \eta_B =\frac{n_B}{n_{\gamma}} = (6.1 \pm 0.2)\times 10^{-10}.
\end{equation}
Interestingly, the big-bang nucleosynthesis is completely consistent
with this determination.

To produce this asymmetry through leptogenesis, several
considerations have to be addressed. First, what is the number of
right-handed neutrinos decaying out of thermal equilibrium? The
answer to this question is in principle depends on the thermal
history of the right-handed neutrinos. In our model, it turns out
that this dependence is rather weak because of the strong washout.
Second, what is the lepton density generated from the right-handed
neutrino decay? This, of course, is related to the CP asymmetry of
the decay which depends on complex phases in the Yukawa
interactions. Third, some of the generated lepton density gets
washed out by inverse-decay processes and scattering. This effect
can be rather important, particularly in the so-called strong
washout region. Finally, one must calculate the percentage of lepton
number density converted into the baryon number density through the
electroweak sphaleron process. The answers to some of the questions
are less model-dependent and are standard in the literature
\cite{Giudice}. Here we focus on the parts depending on particular
models for the right-handed neutrinos.

The density of leptons from right-handed neutrino decays is
\begin{equation}
n_L = \frac{3\zeta(3)g_N T^3}{4\pi^2}\sum_{i=1}^3 \kappa_i
\epsilon_i \label{lepton} \ ,
\end{equation}
where the first factor is the thermal density of a relativistic
fermion with $g_N=2$ and the sum is over the number of
right-handed neutrinos. The $\epsilon_i$ is the decay CP asymmetry
of the $i$-th right-handed neutrino; $\kappa_i$ is the
corresponding efficiency factor, taking into account the fraction
of out-of-equilibrium decays and the washout effect. Both factors
depend on the effective mass defined as
\begin{equation} \widetilde{m}_i  =  \frac{(M_{\nu_D}^{'\dag}
M'_{\nu_{D}} )_{ii}}{M_i} \ .
\end{equation}
where $M'$ denotes the Dirac neutrino mass in the basis in which
the right-handed neutrino matrix is diagonal, $M_{\nu}' = M_{\nu}
U^*$. If $U$ diagonalizes $M_{\nu_R}$, then
\begin{equation}
M_{\nu_{R}} = U \hat{M} U^T = U \begin{pmatrix} M_1 & 0 & 0 \\
0 & M_2 & 0 \\ 0 & 0 & M_3
\end{pmatrix} U^T, \,\,\,\,
\end{equation}
and the right-handed neutrino mass eigenstates are $\chi_i = \sum_f
U_{fi}\chi_f$, where $\chi_f$ is the family basis. Reversing this
relation, one has $\chi_f = \sum_i U_{fi}^* \chi_i$. Thus if the
right-handed field $\chi_f$ is replaced by $\chi_i$ in the Yukawa
coupling, the Yukawa matrix is multiplied on the right by $U^*$.
Hence the above mass relation follows.

The lepton number is converted into the baryon number through the
$B-L$ conserving electroweak sphaleron effect \cite{HT},
\begin{equation}\label{}
n_B = -\frac{8N_G + 4N_H}{22N_G + 13N_H} n_L \ ,
\end{equation}
where $N_G=3$ is the number of fermion families and $N_H$ is the
number of Higgs doublets. In the standard model $N=3$ and $N_H=1$.

The photon density can be calculated from the entropy density $s =
\frac{2}{45} g_* \pi^2 T^3$, where $g_*$ is the effective number of
degrees of freedom, through the relation
\begin{equation}
  s= \frac{4}{3}\frac{\pi^2}{30}\left(2 + \frac{21}{11}\right)
\frac{\pi^2}{2\zeta(3)} n_{\gamma}
\end{equation}
where the second factor takes into account the neutrino
contribution. Ignoring the lightest right-handed neutrino
contribution, $g_*$ is 106.75 in the standard model.

The final ratio of baryon to photon number density through
leptogenesis is
\begin{equation}\label{}
\eta_B = \frac{n_B}{n_{\gamma}} = -\frac{602}{53009} \sum_i \kappa_i
\epsilon_i = - 0.0114 \sum_i \kappa_i\epsilon_i \ .
\end{equation}
Now we turn to the decay asymmetry and efficiency factors.

\subsection{The CP Asymmetry from Right-Handed Neutrino Decay}

The right-handed neutrinos are assumed to be CP eigenstates in the
absence of the Yukawa type of weak interactions. In the presence of
the interactions, they can decay into both left-handed leptons
(neutrino and charged leptons) plus Higgs bosons and right-handed
antileptons plus Higgs bosons. In the leading order, the decay rate
is
\begin{equation}
     \Gamma_i = \frac{1}{8\pi}({Y'}^\dagger Y')_{ii}M_i \ ,
\end{equation}
where again, $Y'=YU^*$ is the Yukawa matrix in the basis where the
right-handed neutrinos are in mass eigenstates.


At next-to-leading order, the decay rates into leptons and
antileptons are different due to the complex phases in the Yukawa
couplings. The decay CP asymmetry is defined as
\begin{equation}
   \epsilon_i
     = \frac{\Gamma(N_i\rightarrow l_jH)-\Gamma(N_i\rightarrow \bar
l_jH^\dagger)
     }{\Gamma(N_i\rightarrow l_jH) + \Gamma(N_i\rightarrow \bar
l_jH^\dagger)
     } \ .
\end{equation}
In one-loop approximation, one finds,
\begin{equation}
    \epsilon_i
      = \frac{1}{8\pi} \sum_{j\ne i}  F\left(\frac{M_j^2}{M_i^2}\right)
            \frac{{\rm Im}[({Y'}^\dagger Y')^2_{ij}]}{({Y'}^\dagger
Y')_{ii}} \ ,
\label{asymm}
\end{equation}
where the decay function is given by \cite{CRFM}
\begin{equation}
   F(x) = \sqrt{x}\left[\frac{1}{1-x}
    + 1-(1+x) \ln\frac{1+x}{x}\right] \ .
\end{equation}
In the limit of large $x$, this become $-3/2\sqrt{x}$. The first
term in $F$ is singular when two right-handed neutrinos become
degenerate in mass, in which case, one must resum the self-energy
corrections which lead to the so-called resonant leptogenesis.


To get a non-zero CP asymmetry, one needs to have complex phases
in the mass matrices. In the model presented in the last section,
we have assumed the right-handed neutrino mass matrix is real. Now
we relax this assumption, and choose the simplest possibility that
the $a$ and $b$ parameters have phases $\alpha$ and $\beta$,
respectively. Thus the new mass matrix becomes,
\begin{equation}
M_{\nu_R}=\begin{pmatrix} ae^{i\alpha} & 0 & 0 \\
0 & be^{i\beta} & 0 \\ 0 & 0 & 1
\end{pmatrix}\Lambda_R \ .
\end{equation}
We refit the parameters $a$, $b$, $\alpha$ and $\beta$ to reproduce
the light-neutrino masses and mixing. Clearly, there is no unique
set of parameters that are capable of reproducing the phenomenology.
One particular set of parameters we use here is
\begin{eqnarray}
  a&=&0.00129\ , ~~~~~~ b=0.00198 \ , \nonumber \\
  \alpha &=& -1.808 \ , ~~~~~ \beta = -3.210 \ .
\end{eqnarray}
Thus the masses of three right-handed neutrinos are then
\begin{equation}
   M_1 = 2.27 \times 10^{10}~{\rm GeV}\ , ~~~~  M_2 = 3.61 \times 10^{10} {\rm
GeV}\ ,
     ~~~~ M_3 = 1.85 \times 10^{13}~ {\rm GeV}\ .
\end{equation}
We see that $M_1$ and $M_2$ are fairly close to each other, with
$\delta =(M_2-M_1)/M_1 = 0.59$. The Yukawa matrix in the basis in
which the right-handed neutrino mass matrix is diagonal and real
looks like
\begin{equation}
Y_{ij}'=\begin{pmatrix} 4.8\times 10^{-6}\exp(0.9i) & 0 & 0.013 \\
0 & 0 & -0.022 \\ 0.0046\exp(0.9i) & -0.022\exp(1.6i) & 0.472\ ,
\end{pmatrix}
\end{equation}
where the signs of the phases are chosen to reproduce the right sign
for the CP asymmetry.

Plugging the Yukawa matrix and mass ratios, we find the following CP
asymmetries,
\begin{eqnarray}
  \epsilon_1 &=& -0.92 \times 10^{-5} \ , \nonumber \\
  \epsilon_2 &=& -0.24\times 10^{-5} \ .
\end{eqnarray}
Here we have also shown the CP asymmetry from the second
right-handed neutrino because its mass is close to the first one and
is potentially important for leptogenesis. The result for
$\epsilon_1$ exceeds slightly the bound derived by Davidson and
Ibarra \cite{DI} because the masses are not so hierarchical.

\subsection{Effective Out-of-Equilibrium Decays}

In our model, $M_1$ is close to $M_2$, and $M_3$ is much heavier.
Thus, it is a good approximation to neglect the CP asymmetries and
lepton number generated from the heaviest right-handed neutrinos
(those with mass $M_3$). However, since $\delta = (M_2-M_1)/M_1$ is
less than 1, one has to consider the full decay and washout effects
from the two light right-handed neutrinos.

The efficiency factor can be calculated by solving the Boltzmann
equation for the right-handed neutrinos and lepton densities. The
result depends on the effective mass $\widetilde m_i$. In the
present case, we find,
\begin{equation}
   \widetilde m_1 = 29.1~ {\rm meV}\ , ~~~~~ \widetilde m_2 = 406~ {\rm
meV}\ ,
\end{equation}
The effective masses determine the so-called decay parameters $
      K_i = \widetilde m_i/m^* $
where $m^* = 16 \pi^{5/2} \sqrt{g^*} v^2/(3\sqrt{5}M_{pl}) = 1.08
\times 10^{-3} $eV. In our case
\begin{equation}
    K_1 = 27.0 \ , ~~~~~~~ K_2 = 376.2\ .
\end{equation}
Since $K_i\gg 1$, we are in the so-called strong washout region. In
this region, the effective factor has little dependence on the
thermal history of the right-handed neutrinos. One can assume for
instance that they are not present in the beginning but are produced
purely by the inverse scattering process.

Since the $M_1$ and $M_2$ are close to each other, one expects that
the existence of $N_2$ will strongly modify the washout of $N_1$.
This situation has been discussed recently in Ref \cite{BD}, where
analytical formulas have been derived from numerical solutions of
the Boltzmann equations,
\begin{equation}
\kappa_1  = \frac{2}{z_B(K_1 + K_2^{(1-\delta)^3}) \cdot (K_1 +
K_2^{(1-\delta)})}\ ,
\end{equation}
\begin{equation}
\kappa_2  = \frac{\left[1+2 \ln\left| \frac{1+\delta}{1-\delta}
\right| \right]^2}{z_B(K_2 + K_1^{(1-\delta)^3}) \cdot (K_2 +
K_1^{(1-\delta)})} e^{-\frac{8\pi}{3}K_1 \left(
\frac{\delta}{1+\delta} \right)^{2.1}} ,
\end{equation}
where $z_B=M_1/T_B$ is the inverse temperature at which the washout
effects are minimized and $\kappa_2$ is valid when $\delta < 1$
\cite{Buchmuller:2004nz}. Plugging in the parameters, we find,
\begin{equation}
    \kappa_1 = 6.8 \times 10^{-3}\ , ~~~~~~ \kappa_2 = 1.3 \times
    10^{-4} \ .
\end{equation}
Thus, because $K_2\gg K_1$, one has $\kappa_1\gg\kappa_2$.
Therefore, the number of out-of-equilibrium decays from $N_2$ is
more than an order of magnitude smaller.


Putting everything together, the baryon asymmetry in our model is
\begin{equation}
\eta_B = - 0.96 \times 10^{-2} \sum_i \kappa_i \epsilon_i = 6.0
\times 10^{-10} \ ,
\end{equation}
which is close to the observation data.

We have also numerically solved the Boltzmann equations to check the
accuracy of the above formula,
\begin{equation}
\frac{d n_{N_i}}{d z}  = -  (D_i +S) \left(  n_{N_i} - n^{\rm
eq}_{N_i} \right)\ ,
\end{equation}
\begin{equation}
\frac{d n_{\mathcal {L}}}{d z}  = - \epsilon_{N_1} D_1 \left(
n_{N_1} - n^{\rm eq}_{N_1} \right) - \epsilon_{N_2} D_2 \left(
n_{N_2} - n^{\rm eq}_{N_2} \right) - W N_\mathcal {L} \ ,
\end{equation}
where $z = M_1 /T$, $n_{N_i}$ is in unit of $n_{\gamma}$ in a
co-moving volume. $D_i$ is the decay width measured in Hubble
expansion rate $H$.
\begin{equation}
D_1 = K_1 z \frac{{\cal K}_1(z)}{{\cal
K}_2(z)},\,\,\,\,\,\,\,\,\,\,\,\,D_2 = K_2 z \frac{ {\cal
K}_1((1+\delta) z)}{{\cal K}_2((1+\delta) z)} \ ,
\end{equation}
where ${\cal K}_{1,2}$ are modified Bessel functions. $S$ is the
scattering rate of the right-handed neutrinos off the Higgs bosons
and gauge bosons. The washout rate $W$ depends on the inverse decay,
right-handed neutrino scattering, and processes involving the
right-handed neutrino in the intermediate state (the $\Delta L=2$
process). Since we are in the strong washout regime, only inverse
decay itself can bring the right-handed neutrinos to thermal
equilibrium, so we may neglect scattering and consider only decays
and inverse decays,
\begin{equation}
W =\frac{1}{4} z^3 \left[ K_1 \cdot {\cal K}_1(z)+ K_2 \cdot
(1+\delta)^4 {\cal K}_1((1+\delta)z) \right] \ .
\end{equation}
The final baryon asymmetry can be calculated via
\begin{equation}
\kappa_i = - \int^{\infty}_{z_i} dz' \frac{d N_i}{dz'} e^{-
\int_{z'}^{\infty} dz'' W(z'')} \ .
\end{equation}
From Eq. (29), we solve the time evolution of the $N_i$
distribution.

\begin{figure}[t]
\begin{center}
\mbox{
\begin{picture}(0,150)(60,0)
\put(-80,-10){\insertfig{9}{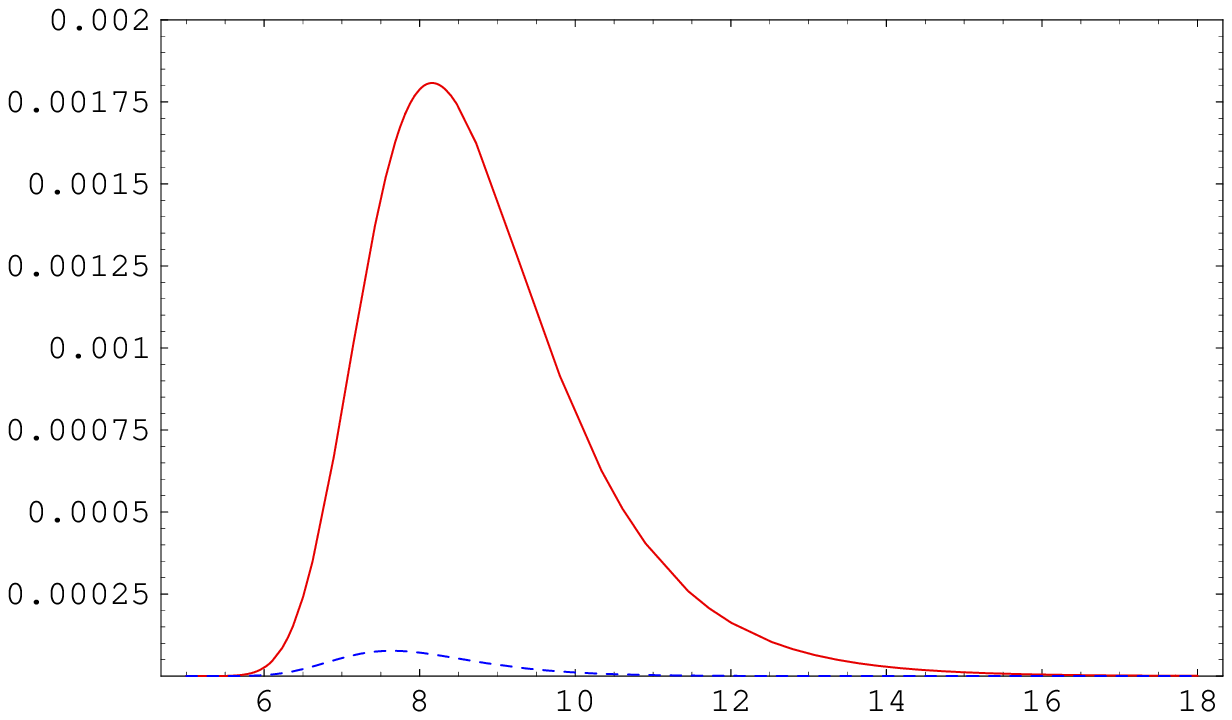}}
\put(-120,75){$d\kappa/dz$} \put(60,-20){$z$}
\end{picture}
}
\end{center}
\caption{\label{fig1} (color online) The efficiency rate for
out-of-equilibrium decays of the lightest (solid line) and the
next-to-lightest (dashed line) right-handed neutrinos as a function
of inverse temperature $z=M_1/T$.}
\end{figure}

It was pointed out \cite{BD} that the main contribution to
$\kappa_i$ comes from a Gaussian-like peak of the integrand around
$z_{B}$. Integrands of both $\kappa_1$ and $\kappa_2$ are depicted
in Fig. 1. We find there is such a peak at around $z_{B}=8 > z^{\rm
eq}$ ($z^{\rm eq}$ represents the time that distribution of the
$i$-th RH neutrino equals the equilibrium distribution and then
follows closely to it in strong washout regime). The asymmetries
generated before thermalization of $N_i$ are negligible, since they
experiences longer washout. The dashed line, which shows the
integrand for $\kappa_2$, is much smaller than the solid line for
$\kappa_1$. So in our model, we can basically neglect the effect of
$N_2$. The numerical results are
\begin{equation}
\kappa_1 = 5.9\times 10^{-3}, \,\,\,\,\, \kappa_2 = 1.4\times
10^{-4} \ .
\end{equation}
Compared with Eq. (27), we find that the numerical solution and the
analytic approximation are reasonably close.


\subsection{Adding Supersymmetry}

In the presence of supersymmetry, for example in minimal
supersymmetric standard model (MSSM), the above leptogensis
calculation must be modified in several ways.

First, the lepton number density now must include the contribution
from the decay of right-handed sneutrinos,
\begin{equation}
n_L = \frac{3\zeta(3)g_N T^3}{4\pi}\sum_{i=1}^3 (\kappa_i \epsilon_i
+  \widetilde \kappa_i \widetilde \epsilon_i) \label{lepton} \ ,
\end{equation}
where, $\widetilde \kappa_i$ and $\widetilde \epsilon_i$ are the
efficient factor and decay asymmetries of the sneutrinos. Because of
supersymmetry, the second term in the sum is the same as the first
term.

Second, when the lepton number is converted into the baryon number
through the sphaleron process, one has now $n_B = -8/23n_L$.

Third, the entropy density $s = \frac{2}{45} g^* \pi^2 T^3$ now has
an effective $g_*$ factor 228.75 in MSSM.

Combining the above, one has
\begin{equation}
    \eta_B = -\frac{344}{77165}\sum_i (\epsilon_i \kappa_i
      + \widetilde \epsilon_i \widetilde \kappa_i)
        = -4.46\times 10^{-3} \sum_i (\epsilon_i \kappa_i
      + \widetilde \epsilon_i \widetilde \kappa_i) \ ,
\end{equation}
which has a coefficient roughly a factor of 2 smaller.

Now consider the decay asymmetry. Equation (\ref{asymm}) for the
right-handed neutrino still applies, except now we have to take into
account the sneutrino intermediate state contribution, the decay
function becomes,
\begin{equation}
   F(x) = -\sqrt{x}\left[\frac{2}{x-1}
    + \ln\frac{1+x}{x}\right] \ .
\end{equation}
In the limit of large-$x$, the above becomes $3/\sqrt{x}$. Thus
the asymmetry is a factor of 2 bigger. In our model, this is
roughly the case:
\begin{equation}
  \epsilon_1 = -1.78 \times 10^{-5}, ~~~~~ \epsilon_2 = - 4.9\times
  10^{-6} \ .
\end{equation}
The decay asymmetry for the sneutrinos $\widetilde \epsilon_i$ are
the same as $\epsilon_i$ because of supersymmetry.

To get the efficiency factors, we turn to the Boltzmann equations.
Not only do we now have an equation for the right-handed neutrinos,
but also for the sneutrinos. Ignoring the effects from scattering
and taking into account the decay and inverse decay, we have
\begin{eqnarray}
\frac{d n_{N_i}}{d z} &=& -  D_i  \left(  n_{N_i} - n^{\rm eq}_{N_i}
\right) \ ,  \nonumber \\
\frac{d n_{\widetilde N_i}}{d z} &=& -  \widetilde D_i  \left(
n_{\widetilde N_i} - n^{\rm eq}_{N_i}
\right)  \ , \nonumber \\
\frac{d n_{\mathcal {L}}}{d z}  &=& - \epsilon_{N_i} D_i \left(
n_{N_i} - n^{\rm eq}_{N_i} \right) - \epsilon_{\widetilde N_i}
\widetilde D_i \left( n_{\widetilde N_i} - n^{\rm eq}_{N_i} \right)
- W n_{\mathcal {L}} \ ,
\end{eqnarray}
where $n_{\widetilde N_i}$ stands for the density of sneutrinos.
The above equations are similar to the
non-supersymmetric case, except for the additional contribution from
the sneutrinos. Because of supersymmetry, the latter contribution to
the lepton density is the same as that of the neutrino. However, the
decay width of the particles is enhanced by a factor of 2, which
leads to a factor of 2 larger K-factors,
\begin{equation}
K_1 = 53.9, \,\,\,\,\,\,\,\,\, K_2 = 752.5 \ .
\end{equation}
We numerically solve the Boltzmann equations and find that
\begin{equation}
\kappa_1 = 3.1\times 10^{-3}, \,\,\,\,\, \kappa_2 = 4.8\times
10^{-5} \ .
\end{equation}
So $\kappa_i$ are about half of what we find in the
non-supersymmetric case.

Tallying all the changes above, we find the final baryon asymmetry
\begin{equation}
  \eta_B = 4.9 \times 10^{-10}  \ ,
\end{equation}
which is slightly smaller than that of the non-supersymmetric case.
But the difference is well within theoretical uncertainties.

In this model, the lightest right-handed neutrino can be produced
thermally if the temperature of the universe after the inflaton
decays is higher than $10^{10}$ GeV. However, if one takes the
cosmological gravitino problem seriously, there is an upper bound
on the reheating temperature $T_R \leq 10^{6}-10^{8}$ GeV when the
gravitino mass is in the range $100$ GeV $\leq m_{3/2}\leq 1$ TeV
\cite{KKM}. In this case the right-handed neutrinos will be
produced non-thermally from the decay of the inflaton. For
example, one may consider the superpotential \cite{FTT}
\begin{equation}
W_{\phi N} = \frac{1}{2}m_{\phi}\phi^2 + g\phi NN \ ,
\end{equation}
where $\phi$ is the inflaton field with mass $m_{\phi} > 2M_1$  and
$g$ is a dimensionless coupling. The reheating temperature is given
by
\begin{equation}
T_R \simeq |g|\sqrt{m_{\phi}M_{pl}} \ .
\end{equation}
Assuming that the branching ratio of the inflaton decay is of
order 1, the produced baryon asymmetry is given by
\begin{equation}
\eta_B \simeq 10^{-10}\;\frac{\epsilon}{2\times10^{-5}} \;
\frac{T_R}{10^6 \rm GeV}\frac{5\times 10^{10}\rm GeV}{m_{\phi}} \ .
\end{equation}
Taking the reheating temperature $T_R \sim 10^{7}$ GeV and $m_{\phi}
\sim 2M_1$ one can still obtain the desired baryon asymmetry of the
universe.

Having shown that our model can generate enough CP violation at high
energy for leptogenesis, it is interesting to calculate the size of
the CP violation at low energy. The low energy CP violation is
encoded into one Dirac CP phase $\delta_{CP}$, which is multiplied
by $\rm{sin}\theta_{13}$ in the standard convention, and two
Majorana phases $\phi_1$ and $\phi_2$, which appear in the form
${\rm diag}(e^{i\phi_1},e^{i\phi_1},1)$ in the PMNS matrix. It has
been shown by Branco, Morozumi, Nobre and Rebelo in \cite{branco}
that there is no model-independent relation between the CP violation
at high and low energies. However, we do have predictions of the low
energy CP phases from our model, and it turns out that they are all
small. The scatter plot of these CP phases versus the
$\rm{sin}^2\theta_{13}$ are shown in Fig. 2. Those points are chosen
to satisfy the $3\sigma$ range of neutrino oscillation data: $0.7
\leq \rm{sin^22\theta_{12}} \leq 0.92$; $ \rm{sin^22\theta_{23}}
\geq 0.87$; $\rm{sin^2\theta_{13}} \leq 0.051$;
$7.1\times10^{-5}{\rm eV}^2\leq\Delta
m^2_{\nu12}\leq8.9\times10^{-5}{\rm eV}^2$; $1.4\times10^{-3}{\rm
eV}^2\leq\Delta m^2_{\nu23}\leq3.3\times10^{-3}{\rm eV}^2$. As shown
in these scatter plots, $\delta_{CP}$ is constrained to be around
$3^\circ$, and the $\phi_1$ and $\phi_2$ are constrained to be
within $3$ degree and $5$ degree deviation from $-180^\circ$ and
$90^\circ$, respectively, indicating small CP violations at low
energy. The prediction of $\rm{sin}^22\theta_{13}$ is shown to be
within the range $0.06 \leq \rm{sin^22\theta_{13}} \leq 0.085$.

\begin{figure}[t]
\begin{center}
\mbox{
\begin{picture}(0,90)(60,0)

\put(-155,0){\insertfig{4.8}{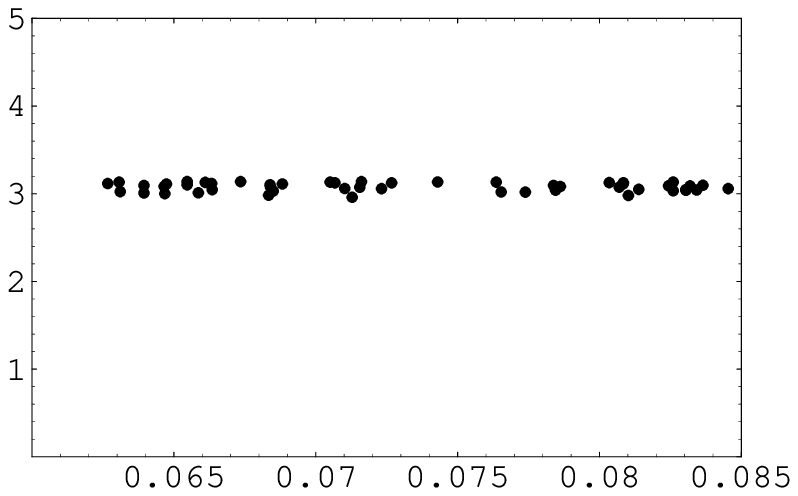}}

\put(-10,0){\insertfig{5.3}{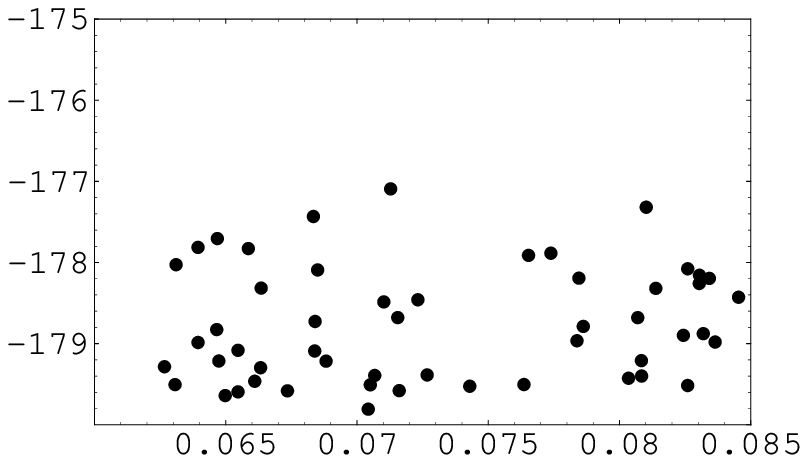}}

\put(150,0){\insertfig{5.0}{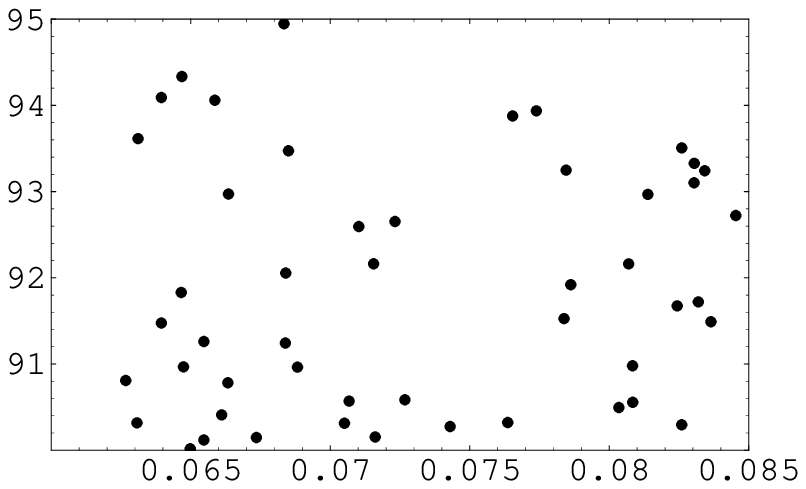}}

\put(-170,75){$\delta_{CP}$} \put(-20,75){$\phi_1$}
\put(140,75){$\phi_2$}

\put(-100,-10){$\rm{sin}^22\theta_{13}$}
\put(55,-10){$\rm{sin}^22\theta_{13}$}
\put(210,-10){$\rm{sin}^22\theta_{13}$}

\end{picture}
}
\end{center}
\caption{The predictions of $\delta_{CP}$, $\phi_1$, and $\phi_2$
plotted against $\rm{sin}^2\theta_{13}$ The points are chosen
according to the requirement of producing enough leptogenesis and
satisfying the $3\sigma$ range of nuetrino oscillation data as
described in the context.}
\end{figure}

\section{Leptogenesis in other SO(10) models}

There are a number of other realistic SO(10) models in the
literature which fit well the quark and charged-lepton properties,
and are consistent with the recent experimental data on the neutrino
mass differences and mixings. However, they provide very different
pictures on leptogensis. The key parameter which controls the main
features of the leptogenesis is the effective mass $\widetilde m_1$:
For a small $\widetilde m_1$, the efficiency factor is large, and
one only needs a moderate value of the decay asymmetry $\epsilon$ to
accomplish leptogenesis. For a large values of $\widetilde m_1$, the
out-of-equilibrium decays are rare, and a successful leptogenesis
requires a large decay asymmetry, which is possible when the masses
of right-handed neutrinos become degenerate (resonant leptogenesis).

In this section, we compare leptogenesis scenarios in different
SO(10) models and comment on their strong and weak points.

\subsection{AB Model}

The AB model \cite{Albright:2001LMA} utilizes Higgs fields $10_H$,
$16_H$, $\overline{16}_H$ and $45_H$. The Dirac mass matrices of
fermions are as follows:
\begin{equation}
M_u = \left(
\begin{array}{ccc}
\eta &  0 & 0 \\
0 & 0 & -\epsilon/3 \\
0 & \epsilon/3 & 1
\end{array}
\right) \Lambda_U\  , \,\,\,\,\,\,\,\, M_d = \left(
\begin{array}{ccc}
0 &  \delta & \delta' e^{i \phi} \\
\delta & 0 & -\epsilon/3 \\
\delta' e^{i \phi} & \sigma+ \epsilon/3 & 1
\end{array}
\right) \Lambda_D \ , \nonumber
\end{equation}
\begin{equation}
M_{\nu_D} = \left(
\begin{array}{ccc}
\eta &  \delta_N & \delta'_N \\
\delta_N & 0 &\epsilon \\
\delta'_N &  -\epsilon & 1
\end{array}
\right) \Lambda_U \ , \,\,\,\,\,\,\,\, M_l = \left(
\begin{array}{ccc}
0 &  \delta & \delta' e^{i \phi} \\
\delta & 0 & \sigma+\epsilon \\
\delta' e^{i \phi} &   -\epsilon & 1
\end{array}
\right) \Lambda_D \ .
\end{equation}
Notice that the mass matrices $M_l$ and $M_d$ are lopsided,
producing a large mixing for the right-handed down quarks as well
as the left-handed charged leptons. Type-one seesaw mechanism is
used to generate the light left-handed neutrino masses. Since the
large-mixing angle MSW solar neutrino solution is preferred by
experiments, the right-handed Majorana mass is constrained to
\cite{Albright:2001LMA}
\begin{equation}
M_{\nu_R} = \left(
\begin{array}{ccc}
c^2 \eta^2 & -b \epsilon \eta & a \eta \\
-b \epsilon \eta & \epsilon^2 & -\epsilon \\
a \eta & -\epsilon & 1
\end{array}
\right) \Lambda_R \ ,
\end{equation}
where $\eta$ and $\epsilon$ are the same parameters as those in
the Dirac mass matrices, and $a$, $b$ and $c$ are additional
parameters of order 1.

A set of parameters which reproduce the quark and charged-lepton
spectra and mixings are,
\begin{equation}
\epsilon = 0.147 \ , \,\,\,\,\,\,\,\,  \eta = 6 \times 10^{-6} \ ,
\nonumber
\end{equation}
\begin{equation}
\delta = 0.00946 \ , \,\,\,\,\,\,\,\,  \delta' = 0.00827 \ ,
\nonumber
\end{equation}
\begin{equation}
\sigma = 1.83 \ , \,\,\,\,\,\,\,\, \phi = 2\pi/3 \ ,  \nonumber
\end{equation}
\begin{equation}
m_U=113~ {\rm GeV}\ , \,\,\,\,\,\,\,\, m_D = 1~{\rm GeV} \ .
\nonumber
\end{equation}
Given the above, additional parameters, $a$, $b$, $c$ and
$\Lambda_R$, can easily be found to fit the neutrino mass
differences and mixings. However, the model usually generates a
very large $\widetilde m_i$, which in turn produces a very large
decay width for the lightest right-handed neutrino. As a
consequence, the efficiency factor $\kappa$ is too small. To
enhance the lepton number production, the masses of the two
lightest right-handed neutrinos are forced to a near degeneracy,
yielding a large resonant decay asymmetry.

In a recent publication, a very extensive search in the parameter
space was conducted to find a viable leptogenesis in the model
\cite{Albright:2005bm}. One of the solutions is described by the
following parameters,
\begin{eqnarray}
&&  \eta=1.1 \times 10^{-5}\ , ~~~~\delta_N=-1.0 \times 10^{-5}\ ,
~~~~\delta'_N=-1.5 \times 10^{-5}\ ,  \nonumber \\
&&  \Lambda_R=2.85 \times 10^{14}~ \rm{GeV}\ ,  \nonumber \\
&& a=c=0.5828i\ ,~~~~~b=1.7670i \ .
\end{eqnarray}
These parameters lead to the following right-handed neutrino masses,
\begin{equation}
  M_1 \sim M_2 = 5.40\times 10^8~{\rm GeV}\ ,  ~~~~~~~~
  M_3 = 2.91 \times 10^{14}~{\rm GeV} \ .
\end{equation}
The $\eta_B$ we calculate from these parameters, however, is $2.6
\times 10^{-6}$, roughly a factor of 2 smaller than that quoted in
Ref. \cite{Albright:2005bm}. The difference comes from the CP
asymmetry of the decay. When the masses of the two right-handed
neutrinos are close, one cannot use the one-loop result in Eq.
(\ref{asymm}) directly. One has to resum the self-energy correction
\cite{Pilaftsis:1997jf} to arrive at
\begin{eqnarray}
\epsilon_1 &\approx& \frac{{\rm Im}[({Y'}^\dagger
Y')^2_{12}]}{8\pi({Y'}^\dagger Y')_{11}} \frac{r_N}{r^2_N\, +\,
[({Y'}^\dagger Y')_{11}/{8\pi}]^2} \ ,\\ \nonumber \epsilon_2
&\approx& \frac{{\rm Im}[({Y'}^\dagger
Y')^2_{21}]}{8\pi({Y'}^\dagger Y')_{22}} \frac{r_N}{r^2_N\, +\,
[({Y'}^\dagger Y')_{22}/{8\pi}]^2}\ .
\end{eqnarray}
where $Y'$ is the $\overline{\nu_L}HN_R$ Yukawa couplings in the
mass eigenstate basis of right-handed neutrinos, and $r_N =
(M^2_{1}-M^2_{2})/(M_{1}M_{2})=-2\delta$ is the degeneracy
parameter.


It is worth pointing out that although the CP asymmetry tends to
be enhanced due to the resonance in the case of two lightest
right-handed neutrinos being quasi-degenerate, the washout effect
is also enlarged in this case. Fortunately, in the present model,
$\widetilde m_2 \sim \widetilde m_1$, so the effect is not
particularly large.
The modified numerical results are listed in Table 1.


\begin{table}
\begin{tabular}{|c|c|c|c|c|c|}\hline
\hline   & BPW   & GMN & JLM&DMM &
AB\\
\hline
$M_1 (GeV)$ &$10^{10}$ & $10^{13}$& $3.77\times 10^{10}$   & $10^{13}$
&$5.4\times10^{8}$\\
$-\epsilon $& $ 2.0 \times 10^{-6}\sin 2 \phi  $ & $ 1.94
\times 10^{-6}  $ & $ 1.0\times 10^{-5}$    &  $10^{-4}\sin 2 \phi$
& $9.4\times
10^{-4} $ \\
$\widetilde{m}_1 (eV)$ & 0.003 & 0.006&  0.026    & 0.1-0.4 & 5.4\\
$\kappa $ & $ 6 \times 10^{-2} $ & $1.2\times 10^{-2}$ & $6.3\times
10^{-3}$      & $10^{-3}$ & $ 1.4
\times 10^{-5}$ \\
$\eta_B $& $ 12 \times 10^{-10} \sin 2 \phi $ & $ 4.97 \times
10^{-10}   $&  $6.2\times 10^{-10}$    &  $10^{-9}\sin{2\phi}$& $
2.6 \times
10^{-10}$  \\
$\sin^2 2\theta_{13}$ & $\leq 0.1 $ & 0.12& 0.06-0.085  &
$0.014-0.048$
&0.01\\
\hline
\end{tabular}
\caption[]{{\baselineskip 15pt \it Predicted mass $M_1$ of the
lightest right-handed neutrino, CP asymmetry $\epsilon$, effective
mass $\widetilde m_1$, efficiency factor $\kappa$ and baryon
asymmetry $\eta_B$, and $\theta_{13}$ in various SO(10) models. The
order is arranged according to the size of $\widetilde m_1$.}
\label{table1} }
\end{table}

\subsection{The BPW Model}

In a model proposed by Babu, Pati, and Wilczek \cite{Babu:1998wi},
the fermion Dirac and Majorana mass matrices have the following
form,
\begin{equation}
M_u = \left(
\begin{array}{ccc}
0 &  \epsilon ' & 0 \\
- \epsilon ' & 0 & \epsilon + \sigma \\
0 & -\epsilon + \sigma & 1
\end{array}
\right) \Lambda_U \ ,\,\,\,\,\,\,\,\, M_d = \left(
\begin{array}{ccc}
0 &  \epsilon ' + \eta ' & 0 \\
- \epsilon ' + \eta ' & 0 & \epsilon + \eta \\
0 & -\epsilon + \eta & 1
\end{array}
\right) \Lambda_D \ , \nonumber
\end{equation}
\begin{equation}
M_{\nu_D} = \left(
\begin{array}{ccc}
0 &   -3 \epsilon ' & 0 \\
3 \epsilon ' & 0 & - 3\epsilon + \sigma \\
0 & 3\epsilon + \sigma & 1
\end{array}
\right) \Lambda_U \ ,\,\,\,\,\,\,\,\, M_l = \left(
\begin{array}{ccc}
0 &  -3\epsilon ' + \eta ' & 0 \\
3 \epsilon ' + \eta ' & 0 & -3\epsilon + \eta \\
0 & 3\epsilon + \eta & 1
\end{array}
\right) \Lambda_D \ , \nonumber
\end{equation}
\begin{equation}
M_{\nu_R} = \left(
\begin{array}{ccc}
x &  0 & z \\
0 & 0 & y \\
z & y & 1
\end{array}
\right) \Lambda_R \ .
\end{equation}
One set of parameters which produces good phenomenology without CP
violation is
\begin{equation}
\sigma = 0.110 \ ,\,\,\,\,\,\,\,\, \eta = 0.151\  ,\,\,\,\,\,\,\,\,
\epsilon = -0.095 \ , \nonumber
\end{equation}
\begin{equation}
\eta' = 4.4 \times 10^{-3}\  ,\,\,\,\,\,\,\,\, \epsilon' = 2 \times
10^{-4}\ ,  \nonumber
\end{equation}
\begin{equation}
x = 10^{-4}\  ,\,\,\,\,\,\,\,\, y = -1/17 \ ,\,\,\,\,\,\,\,\, z =
1/200 \ , \nonumber
\end{equation}
\begin{equation}
\Lambda_U(M_X) = 120~ {\rm GeV}\  ,\,\,\,\,\,\,\,\, \Lambda_D(M_X) =
1.5~ {\rm GeV} \ ,\,\,\,\,\,\,\,\, \Lambda_R = 10^{15}~ {\rm GeV} \
.  \nonumber
\end{equation}
With the above, a number of successful predictions follow, including
the masses for bottom and down quarks, CKM martix elements, and the
atmospheric neutrino oscillation parameters.

However, the solar-neutrino mixing angle from this model comes out
too small: $\sin \theta_{12} = 0.04$, in contradiction with the
experimentally-preferred large-mixing angle solution. To remedy
this, a small intrinsic mass for lefthanded neutrinos is introduced,
which stems from operators $\kappa_{12}16_1 16_2 16_H 16_H 10_H
10_H/M_G^3$. With this modification, the solar mixing angle changes
to $\sin^2 2 \theta_{12} \simeq 0.6$ \cite{pati}.

The result of leptogenesis in this model is summarized under ``BPW"
in Table I. Instead of showing a range of results, as in Ref.
\cite{pati1}, we take $\widetilde m_1=3 ~{\rm meV}$, and the
efficiency factor is then about $6\times 10^{-2}$. The baryon
asymmetry is about $12\sin 2\phi\times 10^{-10}$, where $\phi$ is a
CP-violation phase. With a reasonable choice of $\phi$, the
experimental $\eta_B$ is produced.

Clearly the model does not provide a tight constraint on the
relation between the low-energy neutrino properties and
leptogenesis. Introducing an intrinsic mass term for the light
neutrinos might be physically motivated. However, it dilutes the
relation between $M_R$ and low-energy neutrino observables. In an
extreme case, the constraint on the right-handed neutrino properties
will be lost if the low-energy neutrino mass matrix is entirely
``intrinsic", i.e., of non-see-saw origin. A mild reflection of this
``decoupling" is that $\widetilde m_1$ in this model is particularly
small, which is hard to achieve in a complete first-type see-saw
model which fits the low-energy data. A small $\widetilde m_1$
yields a large efficiency factor which certainly aids the
leptogenesis here.

\subsection{The Minimal  $126$-Higgs Model  }

This model \cite{GMN,bm} (referred to as GMN model in the table)
uses the $126_H$ to break $B-L$ symmetry and the $10_H$, $210_H$ to
break gauge symmetries \cite{GMN}, and allows all couplings to be
complex. $10_H$ and $\overline{126}_H$ are used to give fermion
masses through superpotential
\begin{equation}
W_Y = h_{ji} \psi_i \psi_j H_{10} + f_{ij} \psi_i \psi_j
\Delta_{\overline{126}} \ ,
\end{equation}
So the following mass matrices are obtained:
\begin{equation}
M_u = \bar{h} + \bar{f} \ , ~~~~~~~ M_d = \bar{h} r_1 + \bar{f}r_2 \
, \nonumber
\end{equation}
\begin{equation}
M_l = \bar{h} r_1 - 3 \bar{f}r_2\  , ~~~~~~~ M_{\nu_D} = \bar{h} - 3
\bar{f} \ ,
\end{equation}
where $\bar{h}$ and $\bar{f}$ are matrices
\begin{equation}
\bar{h} = h^* \cos \alpha_u \sin \beta\ , ~~~~~~ \bar{f} = f^* e^{i
\gamma_u} \sin \alpha_u \sin \beta \ , \nonumber
\end{equation}
and the parameters are
\begin{equation}
r_1 = \frac{\cos \alpha_d}{\cos \alpha_u} \cot\beta \ , ~~~~~~ r_2 =
e^{i(\gamma_d-\gamma_u)} \frac{\sin \alpha_d}{\sin \alpha_u}
\cot\beta \ .
\end{equation}
Hence, we can get a sumrule for quark and charged-lepton masses
\begin{equation}
k \frac{M_l}{m_{\tau}} = r \frac{M_d}{m_b} + \frac{M_u}{m_t} \ .
\end{equation}

The seesaw formula in this model is given by $ M_{\nu_L} = f v_l -
M_{\nu_D} ( M_{\nu_R} )^{-1} (M_{\nu_D})^T $ and the calculations
are done assuming that the first term dominates over the second. The
left-handed neutrino Majorana mass matrix satisfies a sumrule
\begin{equation}\label{}
M_{\nu_L} = a ( M_l - M_d )^* \ ,
\end{equation}
The right-handed Majorana neutrino mass matrix is given by
choosing the $B-L$ (seesaw) scale $v_{B-L} = 2 \times 10^{14}~{\rm
GeV}$ and $\gamma_u =0$, $\sin \alpha_u \sim \sin \beta  \sim
O(1)$,
\begin{equation}
M_{\nu_R} = f v_{B-L} = \bar{f}^* \frac{e^{-i \gamma_u} }{\sin
\alpha_u \sin \beta } v_{B-L} \simeq (2 \times 10^{14} {\rm GeV})
\bar{f}^* \ .
\end{equation}
Therefore, the left- and right-handed neutrino Majorana masses have
the same texture, and there are some nontrivial relations between
the low energy phenomena and leptogenesis, when the first term is
assumed to dominate in $M_{\nu_L}$.

We use the following parameters and matrices,
\begin{equation}
k = -0.846\ , \,\,\,\,\,\,\,\,r=-1.846 \ ,
\end{equation}
\begin{equation}
r_1 = 0.0116 \ , \,\,\,\,\,\,\,\, r_2 = 0.00337\ ,
\end{equation}
\begin{equation}
\bar{h}= \frac{r_2 M_u -M_d}{r_2 -r_1} = \left(
\begin{array}{ccc}
0.1768  & -0.0207 +0.0038 i & -0.058+0.097i \\
-0.0207 -0.0038i & 3.49 & -1.332 \\
-0.058+0.097i & -1.332  & 94.7
\end{array}
\right) \ ,
\end{equation}
\begin{equation}
\bar{f} = M_u - \bar{h} = \left(
\begin{array}{ccc}
0.165  & 0.0714 - 0.0133 i & 0.2 - 0.335 i \\
0.0714 + 0.0133 i & -3.161 & 4.596  \\
0.2 + 0.335 i & 4.596  & -12.28
\end{array}
\right) \ .
\end{equation}
The predicted baryon asymmetry through thermal leptogenesis is
listed in Table $1$ as ``GMN".

This model is also characterized by a small $\widetilde m_1$,
which is not constrained by the low-energy neutrino mass spectra
because the first type of see-saw mass contribution is assumed to
be small. Note that the neutrino Dirac mass texture in this model
is completely untested at low energy and can only be effective in
the right-handed neutrino decay process.

\subsection{The DMM Model  }

The DMM model \cite{DMM} is an extension of the GMN model by
enlarging the Higgs sector to include $120(A)$ which gives an
additional contribution to the fermion mass matrices through the
coupling
\begin{equation}
W_{120} =  \frac12 h^\prime_{ij} \psi_i \psi_j D \ ,
\end{equation}
where $h^\prime_{ij}$ is an antisymmetric matrix due to the
$SO(10)$ symmetry. Furthermore, the {\bf 10} and {\bf 126}
couplings are chosen real and the {\bf 120} imaginary by using a
$Z_2$ symmetry. In this case there are six pairs of Higgs
doublets: $\varphi_d = (H^{10}_d, A^{1}_d, A^{2}_d,
\overline\Delta_d, \Delta_d, \Phi_d)$, $\varphi_u = (H_{u}^{10},
D_{u}^1, D_{u}^2, \Delta_u, \overline\Delta_u, \Phi_{u})$, where
superscripts $1$, $2$ of $A_{u,d}$ stand for SU(4) singlet and
adjoint pieces under the $G_{422}=\ $SU(4)$\times $SU(2)$\times
$SU(2) decomposition.

The MSSM Higgs doublets are given by
\begin{eqnarray}
H_d &=& U^*_{1a}(\phi_d)_a\  , \\ \nonumber H_u &=&
V^*_{1a}(\phi_u)a\ ,
\end{eqnarray}
where $a=1,..6$, $U$ and $V$ are unitary matrices which diagonalize
the Higgs mass matrix. The Yukawa coupling matrices for fermions are
given by
\begin{eqnarray}
Y_u &=& \bar h + r_2\bar f + r_3 \bar h^\prime\ , \label{Y_u} \\
Y_d &=& r_1(\bar h + \bar f + \bar h^\prime) \ , \label{Y_d} \\
Y_e &=& r_1(\bar h - 3 \bar f + c_e\bar h^\prime)  \ , \label{Y_e}\\
Y_\nu &=& \bar h - 3 r_2\bar  f + c_\nu \bar h^\prime \ ,
\label{Y_nu}
\end{eqnarray}
where the subscripts $u,d,e,\nu$ denote for up-type quark,
down-type quark, charged-lepton, and Dirac neutrino Yukawa
couplings, respectively, and
\begin{eqnarray}
\bar h &\!\!=&\!\! V_{11} h\ ,\quad \bar f = U_{14}/(\sqrt3\, r_1)
f\ ,\quad
\bar h^\prime = (U_{12} + U_{13}/\sqrt3)/r_1 h^\prime\ ,\\
r_1 &\!\!=&\!\! \frac{U_{11}}{V_{11}}\ ,\quad r_2 = r_1
\frac{V_{15}}{U_{14}}\ , \quad
r_3 = r_1 \frac{V_{12} -V_{13}/\sqrt3}{U_{12} + U_{13}/\sqrt3} \ , \\
c_e &\!\!=&\!\! \frac{U_{12} -\sqrt3 U_{13}}{U_{12} +
U_{13}/\sqrt3}\ ,\quad c_\nu = r_1 \frac{V_{12} +\sqrt3
V_{13}}{U_{12} +
U_{13}/\sqrt3} \ .
\end{eqnarray}
The light neutrino mass is obtained as
\begin{equation}
m_\nu^{\rm light} = M_L - M_\nu^D M_{\nu_R}^{-1} (M_\nu^D)^T\ ,
\end{equation}
where $M_\nu^D = Y_\nu \langle H_u \rangle$, $M_L = 2 \sqrt2 f
\langle \overline\Delta_L \rangle$, and $M_R = 2 \sqrt2 f \langle
\overline\Delta_R \rangle$.

It has been shown \cite{DMM} that the Dirac neutrino coupling matrix
\begin{equation}
\hat Y_{\nu} = \left(\begin{array}{ccc}
0.002 & 0.003\,\exp(-1.54\,i) & 0.0026\,\exp(-0.344\,i) \\
-0.0167 & 0.021\,\exp(-1.53\,i) & 0.025\,\exp(3.37\,i)  \\
-0.229& 0.417\,\exp(4.70\,i)& 0.422\,\exp(-3.019\,i)
\end{array}
\right) \ ,
\end{equation}
gives a good fit to $\Delta m^2_{\rm atm}$, $\Delta m^2_{\rm sol}$
and mixing angles with the lightest right-handed neutrino mass
$M_1\simeq 10^{13}$ GeV. In this model there is a correlation
between $U_{e3}$ and $V_{ub}$ as well as $U_{e3}$ and $\Delta
m^2_{sol}/{\Delta m^2_A}$. The former imposes upper bound  on
$U_{e3}$, while the latter gives a lower bound. The predicted values
of $U_{e3}$, lepton asymmetry and the washout factor are given in
Table $1$.

\section{Conclusion}

In summary, we have discussed leptogenesis in a SUSY SO(10) GUT
model for the fermion masses and mixings, which is developed from
the original lopsided model of Albright and Barr
\cite{Albright:1998vf}. We have done a detailed analysis of the
washout factor and find that the model predicts a value for the
baryon to photon ratio ($\eta_B$) of the universe in good
agreement with observations from WMAP as well as the requirement
of a successful nucleosynthesis. We then compare with the same
predictions for other successful SO(10) models in the literature.

X. Ji and Y. Li are partially supported by the U. S. Department of
Energy via grant DE-FG02-93ER-40762 and by National Natural
Science Foundation of China (NSFC). R. Mohapatra and  S.Nasri are
supported by National Science Foundation (NSF) Grant No.
PHY-0354401.

\end{document}